\documentclass[conference]{IEEEtran}
\IEEEoverridecommandlockouts
\usepackage{cite}
\usepackage{amsmath,amssymb,amsfonts}
\usepackage{algorithmic}
\usepackage{graphicx}
\usepackage{textcomp}
\usepackage{color}
\def\BibTeX{{\rm B\kern-.05em{\sc i\kern-.025em b}\kern-.08em
    T\kern-.1667em\lower.7ex\hbox{E}\kern-.125emX}}

\newtheorem{definition}{Definition}
\newtheorem{theorem}{Theorem}
\newtheorem{lemma}{Lemma}

\begin{document}

\title{Qualitative properties and stability analysis of the mathematical model  for a DC-DC electric circuit\\

\thanks{This work was supported by the Ministry of Science and Education of the Russian Federation (Project Nos. FZZS-2024-0003, 121041300060-4).}
}

\author{\IEEEauthorblockN{E.V. Chistyakova}
\IEEEauthorblockA{\textit{Institute for System Dynamics and Control Theory	} \\
\textit{Siberian Branch, Russian Academy Sciences}\\
Irkutsk, Russia \\
Irkutsk National Research Technical University\\
elena.chistyakova@icc.ru}
\and
\IEEEauthorblockN{D.N. Sidorov}
\IEEEauthorblockA{\textit{Energy Systems Institute} \\
\textit{Siberian Branch, Russian Academy Sciences}\\
Irkutsk, Russia \\
Harbin Institute of Technology, Harbin, PRC\\
dsidorov@isem.irk.ru}
\and
\IEEEauthorblockN{A.V. Domyshev}
\IEEEauthorblockA{\textit{Energy Systems Institute} \\
\textit{Siberian Branch, Russian Academy Sciences}\\
Irkutsk, Russia \\
Irkutsk National Research Technical University\\
domyshev@isem.irk.ru}
\and
\IEEEauthorblockN{V.F. Chistyakov}
\IEEEauthorblockA{\textit{Institute for System Dynamics and Control Theory	} \\
\textit{Siberian Branch, Russian Academy Sciences}\\
Irkutsk, Russia \\
chist@icc.ru}

}

\maketitle

\begin{abstract}
This paper describes a simplified model of an electric circuit with a DC-DC converter and a PID-regulator as a system of integral differential equations with an identically singular matrix multiplying the higher derivative of the desired vector-function. We use theoretical results on integral and differential equations and their systems to prove solvability of such a model and analyze its stability. 
\end{abstract}


\section{Introduction}

\begin{figure}[h]
	\centering
	\includegraphics[scale=0.242]{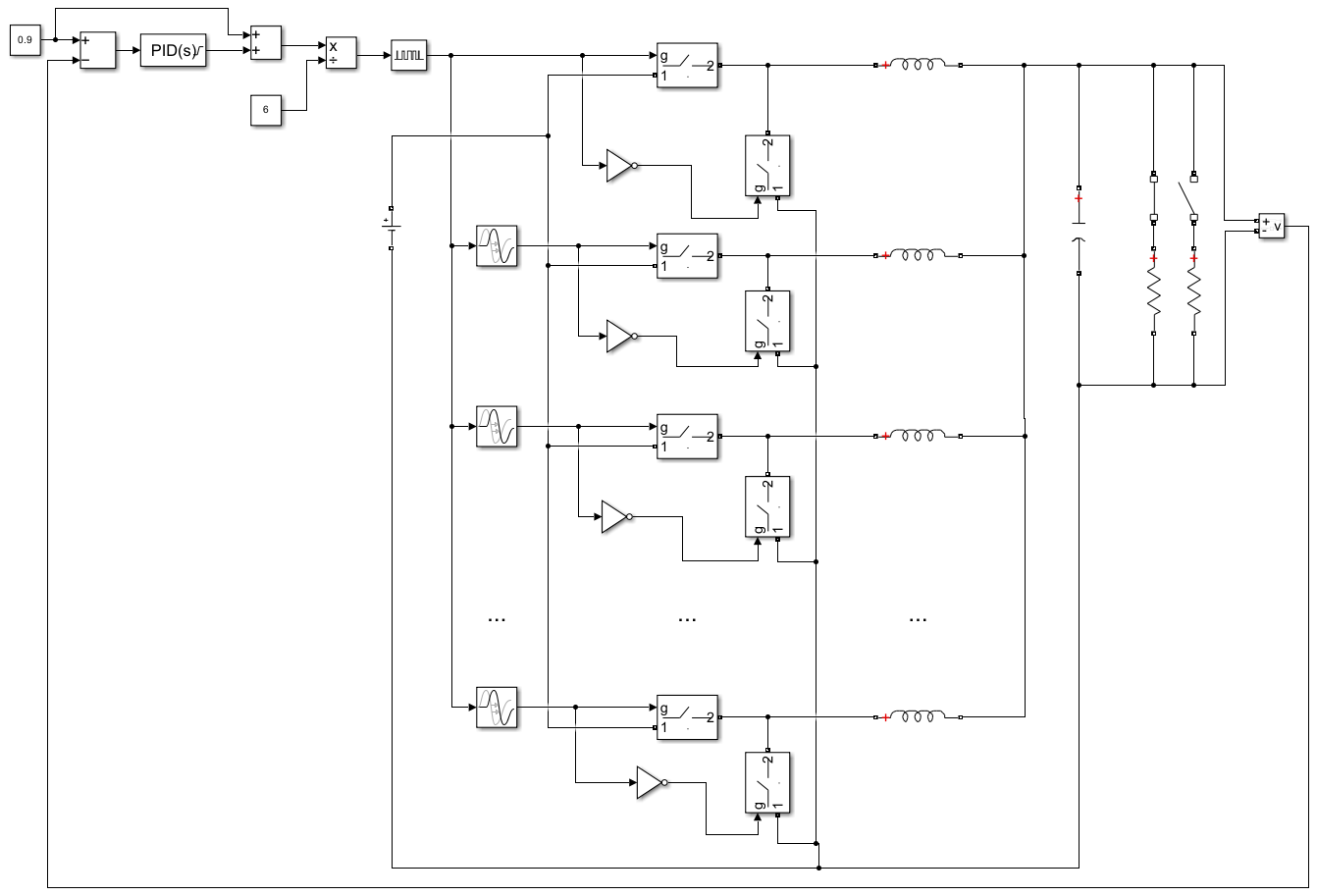}
	\caption{An idealized model of a DC-DC converter}
	\label{fig:DCDCconv}
\end{figure}

DC-DC converters are essential electronic circuits that play a critical role in modern power management systems. Their primary function is to convert the voltage of a direct current (DC) source from one level to another, ensuring stable and efficient power delivery to various electronic devices and systems.
In applications where input voltage levels can fluctuate due to factors such as battery discharging over time or changes in load conditions, DC-DC converters maintain a constant output voltage, providing reliable power to the system's components. DC-DC converters come in various topologies and configurations depending on applications and power requirements \cite{LuoYe2006} . In this paper we consider a mathematical model for an idealized  DC-DC converter (see Fig. \ref{fig:DCDCconv}) in case of big and fast load changes.

The mathematical model that comprises individual equations for the DC-DC converter has the following form:

\begin{equation}\label{modeq1}
	\frac{dI_j}{dt}=\frac{1}{L_j}\bigl( \alpha_j(D_0)U_S-I_jR_L-U_O\bigr), \quad j=1,\ldots N,
\end{equation}
\begin{equation}\label{modeq2}
	\frac{dU_C}{dt}=\frac{1}{C}\Bigl(\sum\limits_{j=1}^N I_j-\frac{U_O}{R_{load}}\Bigr),
\end{equation}
\begin{equation}\label{modeq3}
	U_O=U_C+R_C\Bigl(\sum\limits_{j=1}^{N_f} I_j-\frac{U_O}{R_{load}}\Bigr),
\end{equation}
\begin{equation}\label{modeq4}
	T_d\frac{dU_{ad}}{dt}+U_{ad}=K_d\frac{de}{dt},   
\end{equation}
\begin{equation}\label{modeq5}
	U_{ai}=K_i\int\limits_{t_0}^te(\tau)d\tau,
\end{equation}
\begin{equation}\label{modeq6}
	T_{dd}\frac{dU_{dd}}{dt}+U_{dd}=K_{dd}\frac{d^2e}{dt^2},
\end{equation}
\begin{equation}\label{modeq7}
	U_a=U_{ad}+U_{ai}+K_pe+U_{dd},
\end{equation}
\begin{equation}\label{modeq8}
	e=U_{ref}-U_0,
\end{equation}
\begin{equation}\label{modeq9}
	D_0=\bigl(U_{ref}+U_a\bigr)/U_S,
\end{equation}
\begin{equation}\label{modeq10}
	\alpha_j=
	\left\{ 
	\begin{array}{l}
		0, \quad \mbox{if} \quad (t-\Delta t_j)\, mod \, T>D_0T, \\
		1, \quad \mbox{if} \quad (t-\Delta t_j) \, mod \, T\leq D_0T,
	\end{array} 
	\right.
\end{equation}
In the control cycle, the second derivative of the voltage is calculated using the following expression
\begin{equation}\label{modeq11}
	\frac{d^2U_C}{dt^2}=\frac{1}{C}\Bigl(\sum\limits_{j=1}^{N_f}-\frac{1}{R_{load}}\frac{dU_C}{dt}\Bigr).
\end{equation}
Eqs. (\ref{modeq1})-(\ref{modeq11}) use the following notations:  $N_f$ is number of circuits (phases) of the DC-DC converter, $I_j$ is phase current, $U_o$ is output voltage (actual), $D_0$ is total duty cycle, $U_{ref}$ is the desired output voltage to be maintained, $U_s$ is source voltage, $U_{ad}$ is voltage correction by controller differential gain, $U_{ai}$ is voltage correction by controller integration gain, $U_a$ is total voltage correction, $L_j$ is inductance in each circuit (inductances are assumed to be the same), $R_{load}$ is  load resistance (decreases abruptly -- once during the modeling process,the rate of change of resistance is equal to the current $dR_{load}=-0.9*2000\Sigma/us$, $C$ is a total output capacitance of the DCDC converter, $R_C$ is total series capacitance resistance (ESR), $\Delta t_j$ is deviation of the control start time for each circuit (phase) of the DC-DC converter (uniform filling of the control cycle pulses for the entire DC-DC converter is assumed, so $\Delta t_j=T/N$. During the control process, the duration of the pulses changes, but their relative position along the time axis remains constant), $T$ is control cycle period, $T_d$ is inertia constant of the differential gain, $T_{dd}$ is inertia constant for the second order  derivative gain, $K_d$ is differential gain, $K_p$ is proportional gain, $K_i$ is integration gain, $K_{dd}$
is the correction factor for the second derivative. The controller constants to be optimized are $K_p$, $K_d$, $K_{dd}$, $K_i$, $T_d$, and $T_{dd}$.

\section{Reduction to a system of singular integral differential equations}

Whereas Eqs. (\ref{modeq1})-(\ref{modeq11}) provide a detailed description of the circuit suitable for real-life calculations, it is difficult to analyze qualitative properties of the model, such as existence of solutions, stability, and admissible range of control parameters. Therefore, we suggest that Eqs. (\ref{modeq1})-(\ref{modeq11})  be rewritten in a form of a system of differential, algebraic and integral equations: 

\begin{equation}\label{eqiv1}
	\begin{gathered}
	A_2(\nu ){{d^2z(t)}\over{dt^2}}+A_1(\nu){{dz(t)}\over{dt}}+A_0(\nu)z(t)+\\
	+\int\limits_0^t{\cal K}(t,s,\nu)z(s)ds=F(\nu,z(t)),\ t\in [0,T],\
	\end{gathered}
\end{equation}
where $\nu =(\nu_1, \nu_2,\cdots , \nu_m)$ is a set of scalar parameters that represents values of resistances, inductances, voltages, etc.  $A_2(\nu ),\ A_1(\nu),\ A_0(\nu),\ {\cal K}(t,s)$ are matrices, $z(t)$ is a desirable vector function which components are currents and voltages, and  $T$ is time. The specific feature of the system \eqref{eqiv1} is that
$$
\det A_2(\nu )=0\ \forall \nu,
$$
\begin{equation}\label{eqiv2}
	z(0)=c_j,\ z^{(1)}(0)=q_j,\ z^{(1)}(t)={{dz(t)}/{dt}}.
\end{equation}
where $c_j,q_j$ are given vectors.

For fixed parameters  $\nu =(\nu_1, \nu_2,\cdots , \nu_m),\ \lambda $  and a right-hand part $F(\nu,z(t))$ independent of  $z(t)$, system
 (\ref{eqiv1}) is a special case of a singular system of integral, differential and algebraic equations:
\begin{equation}\label{eqdaeK11}
	\begin{gathered}
	(\bar \Lambda_k+\bar {\cal V})\upsilon:=\sum\limits_{i=0}^k\bar A_i\upsilon^{(i)}(t)+\\ +\int_0 ^t \bar {K}(t-s)\upsilon(s)ds=f(t),\ t\in [0,T]\subset {\bf R}^1,
	\end{gathered}
\end{equation}
where $k=1,2,\ \cdots,\ \bar A_i(t)-(n\times n)$ are constant matrices, $\bar {K}(u)$ is a matrix with real analytical entries in the domain,
$\upsilon \equiv \upsilon(t)$ is a desired vector-function, $f(t)$ is a known vector-function, 
$\upsilon^{(i)}(t)=(d/dt)^i\upsilon(t),$ $\upsilon^{(0)}(t)=\upsilon(t)$.  Most importantly, as above, 
$$
\det \bar A_k=0.
$$

\begin{definition}\label{Def1}
	 If there exists an  $n\times d$-matrix $X_{d}(t)\in{\bf  C}^{k}(T)$, such that any element of the linear solution space of the homogeneous system 	(\ref{eqdaeK11}) is represented as a product 
 $X_{d}(t)c$ on $T$, where $c$ is an arbitrary vector, then we say that this solution space is finite-dimensional $({\rm dim}\; {\rm ker}(\bar \Lambda_{k}+\bar {\cal V})<\infty)$. The smallest possible  $d$ is said to be the dimension of the solution space of system  (\ref{eqdaeK11}).
	
The solution space of the homogeneous system (\ref{eqdaeK11}) is infinite-dimensional  $({\rm dim}\; {\rm ker}(\Lambda_k+{\cal V})=\infty)$, if it comprises an infinite number of linearly independent solutions. 
\end{definition}

\begin{definition}\label{DefOperatorOmega}
	If there exists an operator   
		\begin{equation}\label{operatorOmega}
		\Omega_l=\sum\limits_{j=0}^l\bar L_j(d/dt)^j,\ 
	\end{equation}
	where $\bar L_j$ are constant $n\times n$-matrices, such that  
	\begin{equation}
		\begin{gathered}
		\Omega_l\circ (\Lambda_k+{\cal V})z=(\tilde\Lambda_k+\tilde{\cal V})z=\\
		=\sum\limits_{i=0}^k\tilde A_iz^{(i)}(t)+\int_0 ^t \tilde{K}(t-s)z(s)ds\ \forall z\in {\bf C}^{l+k}(T),
			\end{gathered}
	\end{equation}
		where   $\tilde A_i$ are $n\times n$-matrices from ${\bf C}(T)$, and
	$$
	\det \tilde A_{k}\neq 0,
	$$
	then \eqref{operatorOmega} is said to be the left regularizing operator (LRO) for  the operator $(\Lambda_k+{\cal V})$, and the smallest possible  $l$ is said to be its index.
\end{definition}

\begin{theorem}\label{SPTheorem}
	The solution space of system (\ref{eqdaeK11}) is finite-dimensional $({\rm dim}\; {\rm ker}(\Lambda_k+{\cal V})<\infty)$, if and only if  there exists an LRO for the operator $(\Lambda_k+{\cal V})$. 
\end{theorem}

The sufficiency is straightforward, whereas the necessity is proved similarly to  {\rm \cite{chistMZ2006}}.

\begin{theorem}\label{IndexTheorem}
	If the operator of system (\ref{eqdaeK11}) is index $l$, then the operator  
	$$
	(\bar \Lambda_{k,1}+\bar {\cal V}_1)=\omega_0\circ (\bar \Lambda_k+\bar {\cal V})
	$$
	 is index $l-1$. Here $\omega_0=\begin{pmatrix}P_{1} \cr (d/dt)P_{2}\end{pmatrix}$, and the matrix $P=\begin{pmatrix}P_{1} \cr P_{2}\end{pmatrix}$ is such that  $\det\;P\neq 0,\ P\bar{A}_k =\begin{pmatrix}A_{k,1} \cr 0\end{pmatrix}$, the zero block is of dimension  $(n-r\times n),\ r={\rm rank}\bar{A}_k $. 
	
In Definition \ref{DefOperatorOmega}, we can assume $\Omega_l= \prod_{j=0}^{l-1}\omega_j$, where  $\omega_j$ are corresponding operators. 
\end{theorem}
The proof can be carried out using the same reasoning as in   {\rm \cite{chistMZ2006}}.

\begin{lemma}
	The solution space of the system  $\bar\Lambda_k x:=\sum\limits_{i=0}^k\bar A_ix^{(i)}=0,\ t\in [0,T],$ 
is finite-dimensional (${\rm dim}\; {\rm ker}\ \bar\Lambda _{k}< \infty$), if and only if  $\det\xi(\lambda)=\det\left[\sum\limits_{i=0}^k\lambda^j\bar A_i\right]\not \equiv 0$, where $\lambda$ is a scalar parameter. 
	Here $d=\deg\ \det\xi(\lambda)$,  where $\deg$ stands for the degree of a polynomial, $d$ is an integer parameter from Definition 1. 
		
		If $\det\xi(\lambda)\equiv 0$, then ${\rm dim}\; {\rm ker}\ \bar\Lambda _{k}=\infty$.
	Moreover, the solution space is infinite-dimensional if and only if at least one component of the solution 
	$x\equiv x(t)$ can  be taken as an arbitrary function from  ${\bf  C}^{k}(T)$. 
	
	The operator $\bar\Lambda_k$ has a finite index if and only if  ${\rm dim}\; {\rm ker}\ \bar\Lambda _{k}< \infty$, and  $1\leq l \leq nk$.
\end{lemma}

\begin{lemma}
	The system $\bar\Lambda_k x:=\sum\limits_{i=0}^k\bar A_ix^{(i)}=f(t),\ t\in [0,T],$ is solvable for any  $f(t)\in {\bf C}^{m}(T),\ m\geq l$ if and only if 
	${\rm dim}\; {\rm ker}\ \bar\Lambda _{k}< \infty$, which is equivalent to the condition  $\det\xi(\lambda)=\det\left[\sum\limits_{i=0}^k\lambda^j\bar A_i\right]\not \equiv 0$. 
	
Additionally, 	the general solution to \eqref{eqdaeK11} has the form 
	\begin{equation}\label{eqdaeK12}
		\begin{gathered}
		x(t)=X_{d}(t)c+(\hat{\Lambda}_{l-k}+\hat{{\cal V}})f= \\
		=\sum\limits_{j=0}^k \hat{C}_jf^{(j)}(t)+\int_0 ^t \hat{K}(t-s)f(s)ds,\\ t\in [0,T], \ l\geq k, \ d\geq 0,
		\end{gathered}
	\end{equation} 
	\begin{equation}\label{eqdaeK13}
		\begin{gathered}
		x(t)=X_{d}(t)c+\hat{{\cal V}}f=\\
		=\int_0 ^t \hat{K}(t-s)f(s)ds,\ t\in [0,T], \ l<k, \ d\geq 0,
		\end{gathered}
	\end{equation} 
	\begin{equation}\label{eqdaeK14}
		x(t)=\hat{\Lambda}_{l-k}f
		=\sum\limits_{j=0}^{l-k} \hat{C}_jf^{(j)}(t),\ t\in [0,T], \ d-0, l>k,
	\end{equation} 
	where ${C}_j(t)$ are some $(n\times n)$-constant matrices, $\hat{K}(u)$ is a matrix whose elements are real analytical functions in the domain. 
\end{lemma}

Lemmas 1 and  2 are compiled results from  \cite{Luzin1940} rewritten in a more suitable for the particular applied problem form.

Let us show one simple way of constructing formula (\ref{eqdaeK14}). We have
$$
\begin{gathered}
\det\xi(\lambda)=\det\left[\sum\limits_{i=0}^k\lambda^j\bar A_i\right]={\bf a}_0=const\ \forall \lambda,\\
x(t)={{1}\over{{\bf a}_0}}{\bf A}[(d/dt)]f(t)=\sum\limits_{j=0}^{l-k} \hat{C}_jf^{(j)}(t),
\end{gathered}
$$  
where ${\bf A}[(d/dt)]$ is an algebraic complement matrix to the matrix  $\sum\limits_{i=0}^k(d/dt)^j\bar A_i$ .

For the LRO to the operator $(\bar \Lambda_k+\bar {\cal V})$ to exist, it is not enough to fulfill the condition 
${\rm dim}\; {\rm ker}\ \bar\Lambda _{k}< \infty$. 
Thus, we introduce 
\begin{equation}\label{eqdaeK15}
	\begin{gathered}
(d/dt)^j(\bar \Lambda_k+\bar {\cal V})v:=\\=\sum\limits_{i=0}^{j+k}\check{A}_iv^{(i)}(t)+\int_0 ^t \check{K}_j(t-s)v(s)ds,\ t\in [0,T],
	\end{gathered}
\end{equation}
where 
$$
\begin{gathered}
\check{A}_{j+i}=\bar {A}_i,\ i=\overline{0,k},\ \check{A}_{i}=\check{K}_i(0),\ i=\overline{0,j}\\
\check{K}_j(t-s)={{\partial ^{j}\bar {K}(t-s)}/{\partial t^{j}}}.	
\end{gathered}
$$

 The following statement holds.  

\begin{theorem}
	If  in  (\ref{eqdaeK15}), starting with some $j\geq 0$, the operator 
	$$
	\check{\Lambda}_{j+k}=\sum\limits_{i=0}^{j+k}(d/dt)^i\check{A}_i
	$$ 
	 has index $l_{j}\geq j+k$, then system 
	 (\ref{eqdaeK11}) satisfies the relation $({\rm dim}\; {\rm ker}(\bar \Lambda_{k}+\bar {\cal V})<\infty)$.
	\end{theorem}
The proof follows from (\ref{eqdaeK12}).

The index $l_{j}$ is found using the following algorithm.
If $n(j+k)-d_j$ is a multiple of $n-r$, then  $l_{j}=(n(j+k)-d_j)/(n-r)$, otherwise $l_{j}=[(n(j+k)-d_j)/(n-r)]+1$. Here 
$r={\rm rank}\;\bar {A}_k$, $d_j=\deg\det\left[\sum\limits_{i=0}^{j+k}\lambda^i\check A_i\right]$, $[.]$ stands for the integer part of a number, $d=d_j-nj$, $d$ is a parameter from Definition \ref{Def1}.

Now let is describe a technique for reducing the dimension of system  (\ref{eqiv1}). It is clear that there exist constant matrices 
$P$ and $Q$ with the properties $\det\;PQ\neq 0,$  
\begin{equation}\label{eqdaeK1100}
	\begin{gathered}
P(\bar \Lambda_k+\bar {\cal V})Qw=\\
	=\begin{pmatrix}\sum\limits_{i=0}^k\bar A_{i,1}(Qw(t))^{(i)}+\int_0 ^t \bar {K}_{1}(t-s)(Qw(s))ds\cr 
		\sum\limits_{i=0}^{\kappa }\bar A_{i,2}(Qw(t))^{(i)}+\int_0 ^t \bar {K}_{2}(t-s)(Qw(s))ds\end{pmatrix}-\\
	-\begin{pmatrix}f_1(t)\cr f_2(t)\end{pmatrix},
		\end{gathered}
\end{equation}
where 
$$
\kappa <k,\ \upsilon(t)=Qw(t),\ \bar A_{\kappa ,2}=\begin{pmatrix}\bar A_{\kappa ,21} & \bar A_{\kappa ,22} \end{pmatrix},\ \det\;\bar A_{\kappa ,22} \neq 0.
$$
 Introduce the splitting 
$w(t)=\begin{pmatrix}w_{1}(t)\cr w_{2}(t)\end{pmatrix}$ and transform the second equation from  (\ref{eqdaeK1100}) to the form 
\begin{equation}\label{eqdaeK1101}
	\begin{gathered}
	(\bar \Lambda_{\kappa ,22}+\bar {\cal V}_{22})w_{2}=\\
	=\sum\limits_{i=0}^{\kappa }\bar A_{i,22}w_{2}^{(i)}(t)+\int_0 ^t \bar {K}_{2,22}(t-s)w_{2}(s)ds=\psi(t),
\end{gathered}
\end{equation}
$$
\begin{gathered}
\psi(t)=(\bar \Lambda_{\kappa ,21}+\bar {\cal V}_{21})w_{1}=\\
=-\left[\sum\limits_{i=0}^{\kappa }\bar A_{i,21}w_{1}^{(i)}(t)-\int_0 ^t \bar {K}_{2,21}(t-s)w_{1}(s)ds-f_{2}(t)\right ],
\end{gathered}
$$
where $ \bar {K}_{2}(t-s)=\begin{pmatrix}\bar {K}_{2,21}(t-s) & \bar {K}_{2,22}(t-s) \end{pmatrix}$. Due to $\det\;\bar A_{\kappa ,22} \neq 0$,
Eq. (\ref{eqdaeK1101}) has a solution 
\begin{equation}\label{eqdaeK1102}
	w_{2}(t)=W_{2}(t)c_{2}+\int_0 ^t {\cal K}_{22}(t-s)\psi(s)ds,
\end{equation}
where $c_{2}$ is an arbitrary constant vector, $W_{2}(t)$ is a fundamental matrix, ${\cal K}_{22}(t-s)$ is a resolvent kernel. Substitute the expression for  $w_{2}(t)$ from formula (\ref{eqdaeK1102}) into the first equation of (\ref{eqdaeK1100}) taking
 into account the form of $\psi(t)$. We arrive at the original system of smaller dimension: 
\begin{equation}\label{eqdaeK1103}
	\begin{gathered}
	(\bar \Lambda_{\vartheta}^1+\bar {\cal V}^1)w_{1}=\sum\limits_{i=0}^{\vartheta}\bar A_{i}^1w_{1}^{(i)}(t)+\\
	+\int_0 ^t \bar {K}^1(t-s)w_{1}(s)ds={\cal W}(t)c_{2}+f^{1}(t),
	\end{gathered}
\end{equation}
where $\vartheta\leq k, \bar A_i^1(t)$ are square constant matrices, $\bar {K}^1(u)$ is a matrix whose elements are real analytical functions in the domain, 
${\cal W}(t), \ f^{1}(t)$ are some known matrix and function, whose expressions we do not write down here due to their large sizes. 
Here we take into account that the product of the Volterra operators with convolution kernels is the Volterra operator with a convolution kernel. 

If $\det\bar A_{\vartheta}^1=0$ in (\ref{eqdaeK1103}), then we continue the reduction. Otherwise, we use  (\ref{eqdaeK1102}).  
However, it may happen that the second equation of formula (\ref{eqdaeK1100}) takes the form  $\int_0 ^t \bar {K}_{2}(t-s)(Qw(s))ds-f_{2}(t)=0$.
We differentiate this equation until we obtain a zero matrix  $\bar A_{0,2}$.  And then we use the reduction technique again. If we have $\bar \Lambda_{k}\upsilon=f$, it is much easier to do the reduction since all transformations incorporate only methods of linear algebra. 
\begin{theorem}
	The reduction can performed if and only if there exists an LRO for the operator $(\Lambda_k+{\cal V})$. from system (\ref{eqdaeK11}). 
\end{theorem}

\section{Qualitative properties of the mathematical model for the DC-DC electric circuit}

Using the reduction technique described in Section 3, we transform system  (\ref{eqiv1}) to the following system of equations 
\begin{equation}\label{eqiv3}
	\begin{gathered}
		{{d}\over{dt}}\begin{pmatrix}y_1(t)\cr y_2(t)\end{pmatrix}=
	\begin{pmatrix}p_{11}(\nu ) & p_{12}(\nu )\cr  p_{21}(\nu ) &  p_{22}(\nu  \end{pmatrix}\begin{pmatrix}y_1(t)\cr y_2(t)\end{pmatrix}+\\+\begin{pmatrix}F_{1}(\nu , \alpha(D_{0}))\cr  F_{2}(\nu )\end{pmatrix}, 
	\end{gathered}
\end{equation}
\begin{equation}\label{eqiv4}
	U_{1}=K_{1}y_2(t)+K_{2}{{dy_2(t)}\over{dt}}+\int_0^t{\cal K}(t-\tau)y_2(\tau)d\tau + \varphi(t),
\end{equation} 
\begin{equation}\label{eqiv5}
	\begin{gathered}
		D_{0}=(U_{2}+U_{1})/U_{3},\\ 
		\alpha(D_{0})=\begin{pmatrix}0,\ (t-\Delta t)modT>D_{0}T \cr 1,\ (t-\Delta t)modT\leq D_{0}T\end{pmatrix}
	\end{gathered}
\end{equation}
where $p_{ij}(\nu ),\  F_{1}(\nu , a(D_{0})),\ F_{2}(\nu ),\ \varphi(t),\ {\cal K}(\upsilon)$ are given functions, $K_{1},\ K_{2},\ U_{2},\ U_{3},\ \Delta t$ are given parameters,  $\begin{pmatrix}y_1(t)\cr y_2(t)\end{pmatrix}$ are linear combinations of the components of  $z(t)$,
$$
y_1(t)={\cal I}=\sum\limits_{j=1}^{N_{f}}I_{j},\ y_2(t)=)e=U_{ref}-U_{0},
$$
$$
\begin{gathered}
p_{11}(\nu )=-{{R_{L}}\over{{\cal L}}};\ p_{12}(\nu )={{1}\over{{\cal L}}}N_{f},\\
 p_{21}(\nu )=-{{1}\over{rC}}+{{R_{C}R_{L}}\over{r{\cal L}}},\\ 
p_{22}(\nu )=-\left[{{1}\over{rCR_{load}}}+{{R_{C}}\over{r{\cal L}}}N_{f}\right],\\
r=1+\frac{R_C}{R_{load}},
\end{gathered}
$$
\begin{equation}\label{Ua}
\begin{gathered}
U_{a}=[K_{d}-K_{dd}b]e+K_{dd}{{de}\over{dt}}+\\
+\int_0^t{\cal K}(t-\tau)e(\tau)d\tau + \varphi(t), 
\end{gathered}
\end{equation}
\begin{equation}\label{D0}
\begin{gathered}	
D_{0}=(U_{ref}+U_{a})/U_{S},\\ \alpha(D_{0})=\begin{pmatrix}0,\ (t-\Delta t)modT>D_{0}T \cr 1,\ (t-\Delta t)modT\leq D_{0}T\end{pmatrix}
\end{gathered}
\end{equation}
where
$$
{\cal K}(t-\tau)=K_{i}-\frac{K_{d}}{T_d}a\exp[a(t-\tau)]+\frac{K_{dd}}{T_{dd}}b^2\exp[b(t-\tau)],\ 
$$
$$ 
\begin{gathered}
\varphi(t)=\exp(at)U_{ad}(0)+\exp(bt)U_{dd}(0)-\frac{K_{d}}{T_d}\exp(at)e(0)+\\+\frac{K_{dd}}{T_{dd}}b\exp(bt)e(0)-\frac{K_{dd}}{T_{dd}}\exp(bt){{de}\over{dt}}(0),
\end{gathered}
$$ 
\begin{equation}\label{F1F2}
\begin{gathered}
\tilde F_{1}={{1}\over{{\cal L}}}N_{f}[\alpha(D_{0})U_{S}-U_{ref}],\\ \tilde F_{2}=R_{C}\tilde F_{1}-\left[{{1}\over{CR_{load}}}U_{ref}+r{{d}\over{dt}}U_{ref}\right],
\end{gathered}
\end{equation}
where $a=-\frac{1}{T_d}$, $b=-\frac{1}{T_{dd}}$. 

Now obtain the Lyapunov stability via parameters of the model. The characteristic polynomial takes the form 
$$
\lambda^2 +[p_{11}(\nu )+p_{12}(\nu )]\lambda +[p_{11}(\nu )p_{22}(\nu )]-p_{12}(\nu )p_{21}(\nu )] =0,\ 
$$
and, following the Routh-Hurwitz criterion, we get the stability condition
\begin{equation}\label{StabilityConditions}
	\begin{gathered}
[p_{11}(\nu )+p_{12}(\nu )]>0,\\ [p_{11}(\nu )p_{22}(\nu )]-p_{12}(\nu )p_{21}(\nu )] >0. 
\end{gathered}
\end{equation}

It is important to highlight that the function 
$p_{21}(\nu )]$ may be negative if 
$$
{{1}\over{rC}}<{{R_{C}R_{L}}\over{r{\cal L}}}.
$$ Therefore, the second condition  $[p_{11}(\nu )p_{22}(\nu )]-p_{12}(\nu )p_{21}(\nu )] >0$ might be violated for unsuitable values of
$$
R_{load},\ 
R_{C},\ 
C,\ 
{\cal L.} 
$$  
The parameters $T_{d},\ T_{dd}$ should be taken very small for the function  $\varphi(t)$ to fade out (tend to zero). 

Finally, in the formula for the PID regulator  \eqref{Ua} we have $[K_{d}-K_{dd}b]>0$,  where $b=-1/T_{dd}$.

On the basis of \eqref{F1F2}, the following theorem can  be formulated:

\begin{theorem}
Let the parameters of the system \eqref{eqiv3},\eqref{eqiv4},\eqref{eqiv5} satisfy conditions \eqref{StabilityConditions}. Then, there exists a 
piecewise differentiable solution to   \eqref{eqiv3},\eqref{eqiv4},\eqref{eqiv5} on $[0,T]$. and a constant $\varkappa$ such that 
$$
\Vert \begin{pmatrix}y_1(t)\cr y_2(t)\end{pmatrix}\Vert \leq \varkappa \  \forall T\in\mathbb{R}^1.
$$
\end{theorem}


\vspace{12pt}
\color{red}

\end{document}